\documentclass[prd,superscriptaddress,amsfonts,amssymb,amsmath,showpacs,twocolumn]{revtex4-2}
\usepackage{bm}
\usepackage{amsfonts}
\usepackage{latexsym}
\usepackage{graphicx}
\usepackage{amsmath}
\usepackage{palatino}
\usepackage{mathpazo}
\usepackage{textcomp}
\usepackage{float}
\usepackage{booktabs}
\usepackage{dcolumn}
\usepackage{booktabs}
\usepackage{multirow}
\usepackage{hyperref}
\hypersetup{colorlinks,citecolor=blue}
\usepackage{amsmath}
\usepackage{xcolor}
\usepackage{orcidlink}
\usepackage[caption=false]{subfig}
\usepackage{commath}
\def\jnl@style{\it}
\def\aaref@jnl#1{{\jnl@style#1}}

\def\aaref@jnl#1{{\jnl@style#1}}

\def\aj{\aaref@jnl{AJ}}                   
\def\apj{\aaref@jnl{ApJ}}                 
\def\apjl{\aaref@jnl{ApJ}}                
\def\apjs{\aaref@jnl{ApJS}}               
\def\apss{\aaref@jnl{Ap\&SS}}             
\def\aap{\aaref@jnl{A\&A}}                
\def\aapr{\aaref@jnl{A\&A~Rev.}}          
\def\aaps{\aaref@jnl{A\&AS}}              
\def\mnras{\aaref@jnl{Mon.~Not.~Roy.~Astron.~Soc.}}             
\def\prd{\aaref@jnl{Phys.~Rev.~D}}        
\def\prc{\aaref@jnl{Phys.~Rev.~C}}  
\def\prl{\aaref@jnl{Phys.~Rev.~Lett.}}    
\def\qjras{\aaref@jnl{QJRAS}}             
\def\skytel{\aaref@jnl{S\&T}}             
\def\ssr{\aaref@jnl{Space~Sci.~Rev.}}     
\def\zap{\aaref@jnl{ZAp}}                 
\def\nat{\aaref@jnl{Nature}}              
\def\aplett{\aaref@jnl{Astrophys.~Lett.}} 
\def\apspr{\aaref@jnl{Astrophys.~Space~Phys.~Res.}} 
\def\physrep{\aaref@jnl{Phys.~Rep.}}      
\def\physscr{\aaref@jnl{Phys.~Scr}}       
\def\commat{\aaref@jnl{Comm.~Math.~Phys.}}              
\def\science{\aaref@jnl{Science}}               
\def\cqg{\aaref@jnl{Classical Quant.~Grav.}}            
\def\jpcs{\aaref@jnl{JPCS}}                                     
\def\ijmpd{\aaref@jnl{Int.~J.~Mod.~Phys.~D}}                    
\def\grg{\aaref@jnl{Gen.~Relat.~Gravit.}}               
\def\rpp{\aaref@jnl{Rep.~Prog.~Phys.}}          
\def\npa{\aaref@jnl{Nucl.~Phys.~A}}        
\def\lrr{\aaref@jnl{Living Rev.~Rel.}}                   
\def\jcap{\aaref@jnl{J.~Cosmology Astropart.~Phys.}}    
\def\rmp{\aaref@jnl{Rev.~Mod.~Phys.}}   
\def\epjc{\aaref@jnl{Eur.~Phys.~J.~C}}

\allowdisplaybreaks[1]

\begin{document}

\color{black}       

\title{Bulk viscous fluid in extended symmetric teleparallel gravity}

\author{M. Koussour\orcidlink{0000-0002-4188-0572}}
\email{pr.mouhssine@gmail.com}
\affiliation{Quantum Physics and Magnetism Team, LPMC, Faculty of Science Ben
M'sik,\\
Casablanca Hassan II University,
Morocco.} 

\author{S.H. Shekh\orcidlink{0000-0003-4545-1975}}
\email{da\_salim@rediff.com}
\affiliation{Department of Mathematics. S. P. M. Science and Gilani Arts Commerce
College,\\ Ghatanji, Dist. Yavatmal, Maharashtra-445301, India.}

\author{M. Bennai\orcidlink{0000-0002-7364-5171}}
\email{mdbennai@yahoo.fr}
\affiliation{Quantum Physics and Magnetism Team, LPMC, Faculty of Science Ben
M'sik,\\
Casablanca Hassan II University,
Morocco.} 
\affiliation{Lab of High Energy Physics, Modeling and Simulations, Faculty of
Science,\\
University Mohammed V-Agdal, Rabat, Morocco.}

\author{T. Ouali\orcidlink{0000-0002-0776-1625}}
\email{ouali1962@gmail.com}
\affiliation{Laboratory of physics of matter and radiations,
University Mohammed First,\\ Oujda, Morocco.}

\date{\today}

\begin{abstract}
In this paper, we investigate the existence of bulk viscous FLRW cosmological models in a recently proposed extended symmetric teleparallel
gravity or $f\left( Q,T\right) $ gravity in which $Q$ is the non-metricity
and $T$ is the trace of the energy-momentum tensor. We consider a simple
coupling between matter and non-metricity, specifically, $f\left( Q,T\right)
=\alpha Q^{m+1}+\lambda T$ and $f\left( Q,T\right) =\alpha Q+\lambda T$
where $\alpha $, $\lambda $ and $m$ are free model parameters. The exact
cosmological solutions are found by assuming the scale factor in the form of
the hybrid expansion law. This type of relation generates a time-varying
deceleration parameter with the transition of the Universe from the early
decelerating phase to the present accelerating phase. In the presence of
viscous fluid, we analyze some cosmological parameters of our cosmological
model such as the energy density, bulk viscous pressure, bulk viscous
coefficient, equation of state parameter, and energy conditions. Finally, we
conclude our $f\left( Q,T\right) $\ cosmological models agree with the
recent astronomical observations.
\end{abstract}

\maketitle

\section{Introduction}

Modern cosmology starts its beginnings with the emergence of the theory of
general relativity (GR) at the beginning of the last century, since that
time this branch of physics has experienced great prosperity thanks to the
development of means and tools of observation. On the other hand,
considering the temporal evolution of the Universe, cosmologists argued that
the Universe is in decelerated expansion due to Friedmann's standard
equations based on GR. But at the end of the last century, a cluster of
astronomical observations emerged that showed the opposite to be true,
meaning the Universe is in a phase of accelerated expansion \cite{ref1,
ref2, ref3, ref4}. This cosmic puzzle created a challenge for cosmologists,
which made them wonder about the reasons for this cosmic acceleration.

In the literature, there are many alternatives to GR that attempt to explain
this problem \cite{ref5, ref6, ref7}. The most popular alternative currently
proposed are modified gravity theories such as $f\left( R\right) $ gravity,
where $f\left( R\right) $ is an arbitrary function of the Ricci scalar $R$.
Several models have been studied in framework of $f\left( R\right) $ gravity
in various cosmological contexts \cite{ref8, ref9, ref10}. Next, an
extension of this theory proposed by Harko et al. \cite{ref11} named $%
f\left( R,T\right) $ gravity, where $T$ is the trace of energy-momentum
tensor. For more details on this theory, see our work \cite{ref12, ref13,
ref14, ref15}.

Recently, $f\left( Q\right) $ gravity (or symmetric teleparallel gravity)
appeared, which is a new modified theory of gravity proposed by Jimenez et
al. \cite{ref16}\ where the non-metricity scalar $Q$ drives the
gravitational interaction. This theory is based on Weyl geometry, which is a
generalization of Riemannian geometry that is the mathematical basis of GR.
In Weyl geometry, gravitational effects do not occur because of the change
in direction of a vector in parallel transport, but because of the change in
length of the vector itself. The implications of this theory have been
studied in several papers. The first cosmological solutions in $f\left(
Q\right) $ gravity studied in Ref. \cite{ref17, ref18} while in other works
energy conditions and cosmography in $f\left( Q\right) $ gravity are
discussed in \cite{ref19, ref20}. Quantum cosmology with a power-law model
has been investigated in \cite{ref21}. Cosmological solutions and growth
index of matter perturbations have been evaluated for a polynomial
functional form of $f\left( Q\right) $ gravity in \cite{ref18}. Assuming the
power-law function, the coupling of matter in $f\left( Q\right) $ gravity
analyzed by Harko et al. in \cite{ref23} along with many other works
mentioned in \cite{ref24, ref25, ref26, ref27, ref28}.

Inspired by $f\left( R,T\right) $ gravity, the cosmologist's Xu et al.
proposed an extension of $f\left( Q\right) $ gravity by adding a coupling
between the non-metricity scalar $Q$ and trace of the energy-momentum tensor 
$T$ \cite{ref29}. The Lagrangian of the gravitational field is presumed to
be a general function of both $Q$ and $T$. In addition, the field equations
of this theory are derived from the modified Einstein-Hilbert type
variational principle. Thus, the action for this modified theory of gravity
is given by (in the units $G=c=1$) \cite{ref29}

\begin{equation}
S=\int \left( \frac{1}{16\pi }f(Q,T)+L_{m}\right) \sqrt{-g}d^{4}x,
\label{eqn1}
\end{equation}

Here,\ $f(Q,T)$ is the general function of the non-metricity scalar $Q$ and
trace of energy-momentum tensor $T$, $g$ is the determinant of the metric
tensor $g_{\mu \nu }$ i.e. $g=\det \left( g_{\mu \nu }\right) $, and $L_{m}$
represents the matter Lagrangian density.\ If $L_{m}$ depends only on the
metric components and not on its derivatives, one has, for the
energy-momentum tensor $T_{\mu \nu }$ the following

\begin{equation}
T_{\mu \nu }=\frac{-2}{\sqrt{-g}}\frac{\delta \left( \sqrt{-g}L_{m}\right) }{%
\delta g^{\mu \nu }},  \label{eqn2}
\end{equation}%
so that $T=g^{\mu \nu }T_{\mu \nu }$, and 
\begin{equation}
\theta _{\mu \nu }=g^{\alpha \beta }\frac{\delta T_{\alpha \beta }}{\delta
g^{\mu \nu }}.  \label{eqn3}
\end{equation}

The variation of energy-momentum tensor with respect to the metric tensor $%
g_{\mu \nu }$\ read as, 
\begin{equation}
\frac{\delta g^{\mu \nu }T_{\mu \nu }}{\delta g^{\alpha \beta }}=T_{\mu \nu
}+\theta _{\mu \nu }.  \label{eqn4}
\end{equation}

The non-metricity scalar $Q$ can be given as, 
\begin{equation}
Q\equiv -g^{\mu \nu }(L_{\,\,\,\alpha \mu }^{\beta }L_{\,\,\,\nu \beta
}^{\alpha }-L_{\,\,\,\alpha \beta }^{\beta }L_{\,\,\,\mu \nu }^{\alpha }),
\label{eqn5}
\end{equation}%
where the disformation tensor $L_{\alpha \gamma }^{\beta }$ is given by, 
\begin{equation}
L_{\alpha \gamma }^{\beta }=-\frac{1}{2}g^{\beta \eta }(\nabla _{\gamma
}g_{\alpha \eta }+\nabla _{\alpha }g_{\eta \gamma }-\nabla _{\eta }g_{\alpha
\gamma }).  \label{eqn6}
\end{equation}

Further, the non-metricity tensor $Q_{\gamma \mu \nu }$ is defined as, 
\begin{equation}
Q_{\gamma \mu \nu }=\nabla _{\gamma }g_{\mu \nu },  \label{eqn7}
\end{equation}%
and the trace of the non-metricity tensor is obtained as, 
\begin{equation}
Q_{\beta }=g^{\mu \nu }Q_{\beta \mu \nu }\qquad \widetilde{Q}_{\beta
}=g^{\mu \nu }Q_{\mu \beta \nu }.  \label{eqn8}
\end{equation}

The superpotential of the model is defined as, 
\begin{equation}
P_{\,\,\,\mu \nu }^{\beta }=-\frac{1}{2}L_{\,\,\,\mu \nu }^{\beta }+\frac{1}{%
4}(Q^{\beta }-\widetilde{Q}^{\beta })g_{\mu \nu }-\frac{1}{4}\delta _{(\mu
}^{\beta }Q_{\nu )},  \label{eqn9}
\end{equation}%
and using this definition above, the non-metricity scalar in terms of
superpotential is given as, 
\begin{equation}
Q=-Q_{\beta \mu \nu }P^{\beta \mu \nu }.  \label{eqn10}
\end{equation}

Now, varying the gravitational action (\ref{eqn1}) with respect to metric
tensor $g_{\mu \nu }$, the corresponding fleld equations of $f\left(
Q,T\right) $ gravity is obtained as, 
\begin{widetext}
\begin{equation}
-\frac{2}{\sqrt{-g}}\nabla_{\beta}(f_{Q}\sqrt{-g} P^{\beta}_{\,\,\,\, \mu
\nu}-\frac{1}{2}f g_{\mu \nu}+ f_{T}(T_{\mu \nu}+\Theta_{\mu \nu}) \\
-f_{Q}(P_{\mu \beta \alpha}Q_{\nu}^{\,\,\, \beta \alpha}-2Q^{\beta
\alpha}_{\, \, \, \mu}P_{\beta \alpha\nu})= 8\pi T_{\mu \nu}.  \label{eqn11}
\end{equation}
\end{widetext}In the present paper we use the notations as $f\left(
Q,T\right) \equiv f$, $f_{Q}=\frac{df\left( Q,T\right) }{dQ}$, $f_{T}=\frac{%
df\left( Q,T\right) }{dT}$, and $\nabla _{\beta }$\ denotes the covariant
derivative. From Eq. (\ref{eqn11}) it appears that the field equations of $%
f\left( Q,T\right) $ gravity depends on the tensor $\theta _{\mu \nu }$.
Hence, depending on the nature of the source of matter, several theoretical
models corresponding to different matter sources in $f\left( Q,T\right) $
gravity are possible. Originally, Xu et al. \cite{ref29} obtained three
models using following functional forms of $f\left( Q,T\right) $ as

\begin{itemize}
\item $f(Q,T)=\alpha Q+\lambda T$,

\item $f(Q,T)=\alpha Q^{n+1}+\lambda T$,

\item $f(Q,T)=-\alpha Q-\lambda T^{2}$.
\end{itemize}

where $\alpha $, $\lambda $ and $m$ are constants.

Several authors have explored applications of this theory in many contexts:
Arora et al. \cite{ref30} have recently explored $f(Q,T)$ gravity models
with observational constraints. Yang et al. \cite{ref31} formulated the
geodesic deviation and Raychaudhuri equations in $f(Q,T)$ gravity placed on
the observation that the curvature-matter coupling significantly modifies
the nature of current forces and the equation of motion in the Newtonian
limit. Pati et al. \cite{ref32} considered some rip cosmological models in
the form of $f(Q,T)=aQ^{m}+bT$\ gravity. Also in another work, Pati et al. 
\cite{ref33} studied the dynamical aspects of some accelerating models in $%
f(Q,T)$ gravity using the hybrid scale factor.

Most of the authors have analyzed cosmological models using the perfect
fluid as the content of the Universe to account for various problems in the
scientific domain such as present cosmic acceleration and dark energy. In
line with recent observations, the cosmic acceleration is due to a strange
form of energy wearing negative pressure. Inspired by these observations, in
this article, we will establish a cosmological model without taking dark
energy into account by choosing a further realistic fluid, like a viscous
fluid. Various researchers point out the idea that cosmic viscosity acts as
an alternative to dark energy, which could play an important role in
establishing the accelerated expansion phase of the Universe by devouring
negative effective pressure \cite{ref34}. Diverse research has exhibited
that in the early Universe, matter behaved as a viscous fluid during the
neutrino decoupling in the radiation era \cite{ref35, ref36}. For further
justifications for choosing a viscous fluid instead of a perfect fluid, see 
\cite{ref37}. In addition, many viscous fluid cosmological models have been
considered in the literature. Srivastava and Singh \cite{ref38} evaluated a
new holographic dark energy model within the framework of an FLRW model with
bulk viscous matter content regarding $p_{v}=p-3\zeta H$, where $\zeta $ is
the constant bulk viscosity coefficient and $H$ is the Hubble parameter.
Also, Brevik et al. \cite{ref39} analyzed viscous FLRW cosmology in modified
gravity. Singh and Kumar \cite{ref40}\ created bulk viscosity in $f\left(
R,T\right) $ gravity with the viscous term as $\zeta =\zeta _{0}+\zeta _{1}H$%
, where $\zeta _{0}$ and $\zeta _{1}$ are constants. To get the accelerating
expansion of the Universe, Ren et al. \cite{ref41} supposed the form of the
viscosity coefficient depends on the velocity and acceleration as $\zeta
=\zeta _{0}+\zeta _{1}H+\zeta _{2}\left( \frac{\overset{.}{H}}{H}+H\right) $
where $\zeta _{0}$, $\zeta _{1}$ and $\zeta _{2}$ are constants.

This manuscript is organized as follows: In Sec. \ref{sec2} we derive the
exact solutions of $f\left( Q,T\right) $ gravity for the flat FLRW
space-time. In Secs. \ref{sec3} and \ref{sec4} we analyze the physical
behavior of some cosmological parameters such as the energy density, bulk
viscous pressure, bulk viscous coefficient, equation of state (EoS)
parameter, and energy conditions for both models i.e. $f\left( Q,T\right)
=\alpha Q^{m+1}+\lambda T$ and $f\left( Q,T\right) =\alpha Q+\lambda T$,
respectively. The last Sec. \ref{sec5} is devoted to discuss the results and
conclusion.

\section{Field equations and solutions}

\label{sec2}

Taking into account the spatial isotropy and homogeneity of the Universe, we
consider the following flat FLRW metric for our analysis \cite{ref29} 
\begin{equation}
ds^{2}=-N^{2}\left( t\right) dt^{2}+a^{2}(t)\left[ dr^{2}+r^{2}\left(
d\theta ^{2}+\sin ^{2}\theta d\phi ^{2}\right) \right] ,  \label{eqn13}
\end{equation}%
where $a\left( t\right) $ is the scale factor of the Universe, $N\left(
t\right) $\ is the lapse function considered to be $1$ in the standard case
and ($t$, $r$, $\theta $, $\phi $) are the comoving coordinates. The rates
of expansion and dilation are determined as $H\equiv \frac{\overset{.}{a}}{a}
$ and $T\equiv \frac{\overset{.}{N}}{N}$ respectively. Thus, the
corresponding non-metricity scalar is given by $Q=6\frac{H^{2}}{N^{2}}$,
choosing $N=1$ we have $Q=6H^{2}$. In the study of cosmological
model, for the description of bulk viscosity following two main formalism
are mentioned:\newline
a) The first is the non-casual theory in which the deviation of only
first-order is considered and one can find that the heat flow and viscosity
propagate with infinite speed,\newline
b) The second one is the casual theory which propagates with finite speed.
To analyze the late acceleration of the universe, the casual theory of bulk
viscosity has been used. Cataldo et al. have investigated the late time
acceleration using the casual theory \cite{M}. Basically, they used an
ansatz for the Hubble parameter (inspired by the Eckart theory) and they
have shown the transition of the universe from the big rip singularity to
the phantom behaviour. Hence, here we consider the energy momentum tensor
in the form of bulk viscus fluid as, 
\begin{equation}
T_{\mu \nu }=\left( \rho +\overline{p}\right) u_{\mu }u_{\nu }+\overline{p}%
g_{\mu \nu }.  \label{eqn14}
\end{equation}%
where $u_{\mu }=\left( 1,0,0,0\right) $ is the four velocity vector in
co-moving coordinate system satisfying $u^{\mu }u^{\nu }$ $g_{\mu \nu }=-1$.
Thus, for tensor $\theta _{\mu \nu }$ the expression is obtained as $\theta
_{\mu \nu }=\overline{p}g_{\mu \nu }-2T_{\mu \nu }$, and $\overline{p}$ is
the bulk viscous pressure given by 
\begin{equation}
\overline{p}=p-3\xi H,  \label{eqn15}
\end{equation}%
which satisfies the linear equation of state $p=\gamma \rho $, $0\leq \gamma
\leq 1$, $\xi $ is the bulk viscous coefficient, $H$ is the Hubble
parameter, $p$ is normal pressure and $\rho $ is the energy density of the
Universe.

The trace of energy momentum tensor is given as 
\begin{equation}
T=3\overline{p}-\rho .  \label{eqn16}
\end{equation}

Using the metric (\ref{eqn13}), the Einstein field equations are given as 
\cite{ref29}, 
\begin{equation}
8\pi \rho =\frac{f}{2}-6FH^{2}-\frac{2\widetilde{G}}{1+\widetilde{G}}\left( 
\overset{.}{F}H+F\overset{.}{H}\right) ,  \label{eqn17}
\end{equation}
\begin{equation}
8\pi \overline{p}=-\frac{f}{2}+6FH^{2}+2\left( \overset{.}{F}H+F\overset{.}{H%
}\right) .  \label{eqn18}
\end{equation}%
where $\left( \text{\textperiodcentered }\right) $ dot represents a
derivative with respect to cosmic time $\left( t\right) $. In this case, $%
F\equiv f_{Q}$ and $8\pi \widetilde{G}\equiv f_{T}$ represent
differentiation with respect to $Q$ and $T$ respectively. The evolution of
equation for the Hubble function $H$ can be obtained by combining Eqs. (\ref%
{eqn17}) and (\ref{eqn18}) as, 
\begin{equation}
\overset{.}{H}+\frac{\overset{.}{F}}{F}H=\frac{4\pi }{F}\left( 1+\widetilde{G%
}\right) \left( \rho +\overline{p}\right) .  \label{eqn19}
\end{equation}

From the above equations, there is a system of nonlinear equations with four
unknowns: $H$, $\rho $, $\overline{p}$, $f$. Therefore, to find the exact
solutions and reduce the number of unknowns, additional constraints must be
added. Several researchers in the literature use additional constraints of
the scale factor $a\left( t\right) $ in the form of the power-law and
exponential law to construct a cosmological model that describes the
evolution of the Universe. However, one of the negative aspects of these
models is that they do not take into account the transition of the Universe
from early decelerating to the current accelerating stage, meaning that the
deceleration parameter i.e. $q=-1+\frac{d}{dr}\left( \frac{1}{H}\right) $
remains constant throughout the cosmic evolution. Here, to overcome this
problem, the hybrid expansion law is a good alternative, which is of the
form \cite{ref42,ref42a,ref42b, ref42c} 
\begin{equation}
a=\left[ t^{\eta }\exp \left( \beta t\right) \right] ^{\frac{1}{n}},
\label{eqn20}
\end{equation}%
where $\eta $, $\beta $ and $n$ are positive constants.

From the above equation, the Hubble parameter is obtained as 
\begin{equation}
H=\frac{\beta t+\eta }{nt}.  \label{eqn21}
\end{equation}

Using Eq. (\ref{eqn21}), the deceleration parameter corresponding to the
scale factor (\ref{eqn20}) is given as 
\begin{equation}
q=-1+\frac{d}{dt}\left( \frac{1}{H}\right) =-1+n\eta \left( \beta t+\eta
\right) ^{-2}.  \label{eqn22}
\end{equation}

\begin{figure}[h]
\centerline{\includegraphics[scale=0.65]{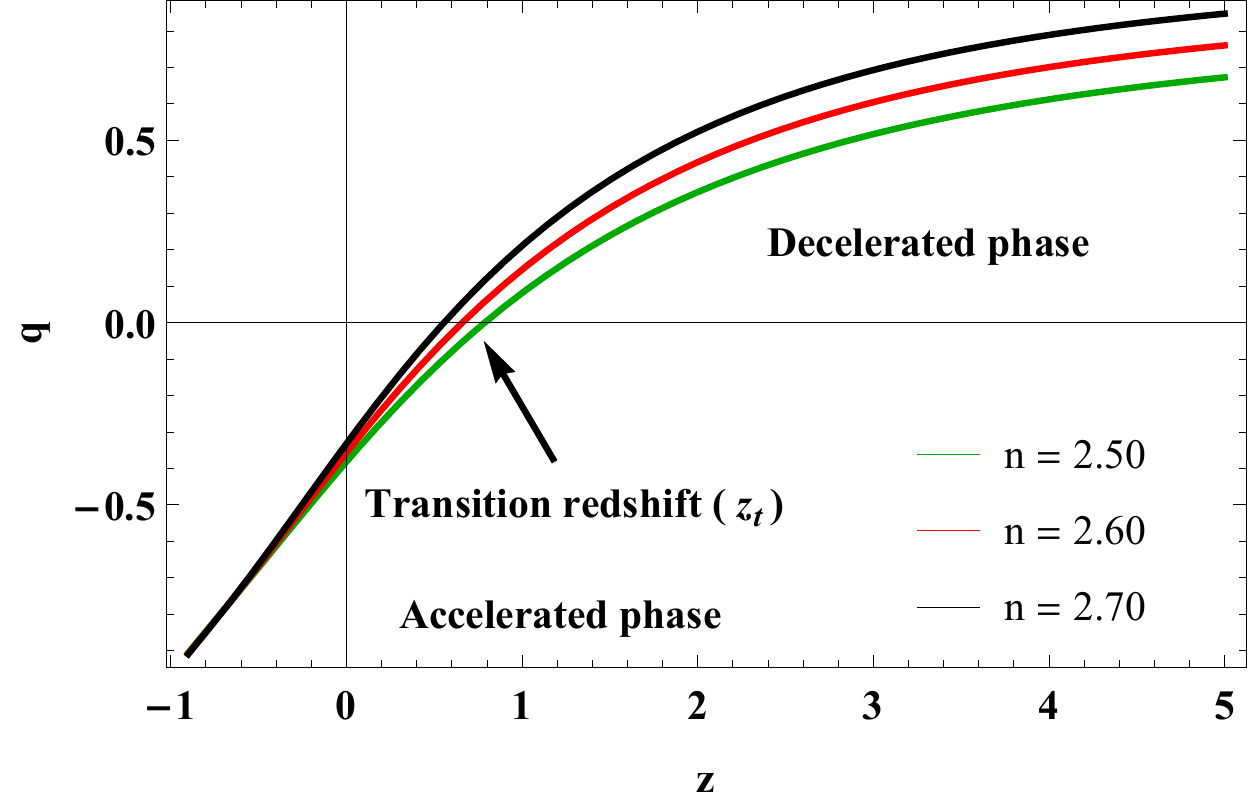}}
\caption{$q$ versus redshift $z$ with $\protect\eta =1.35$ and $\protect%
\beta =2.05$.}
\label{fig_q}
\end{figure}

In cosmology, it is a good idea to write cosmological parameters in terms of
the redshift. Using the relation between the scale factor and redshift of
the Universe as $a\left( t\right) =a_{0}\left( 1+z\right) ^{-1}$ where $%
a_{0}=1$ represents the present value of the scale factor, the time-redshift
relation is given by, 
\begin{equation}
t\left( z\right) =\frac{\eta }{\beta }W\left[ \frac{\beta \left( z+1\right)
^{-\frac{n}{\eta }}}{\eta }\right] .  \label{eqn23}
\end{equation}%
where $W\equiv Lambert$ denotes the Lambert function (also known as "product
logarithm").

The deceleration parameter is a good tool to describe the evolution of the
Universe from the initial deceleration to the current acceleration. If $q>0$
the Universe decelerates, and if $q<0$ then it accelerates. Recent
observations in astronomy such as SNIa (type Ia Supernovae) \cite{ref1, ref2}
and CMB (Cosmic Microwave Background ) anisotropy \cite{ref3, ref4} have
shown that the Universe is already in an acceleration phase and the current
value is in the range $-1\leq q\leq 0$. By using Eqs. (\ref{eqn22})
and (\ref{eqn23}), the variation of the deceleration parameter ($q$) in
terms of the redshift $z$ for the hybrid scale factor is exhibited in Fig. %
\ref{fig_q} for the three representative values of $n$ i.e. $2.50$, $2.60$, $%
2.70$. It is clear that for all $n$ the deceleration parameter is positive
at early phase of the Universe and negative for the late Universe. Hence, it
indicates that the Universe shows a transition from deceleration to
acceleration. Also, the transition from the early deceleration phase to the
current accelerated phase is done with a certain redshift, called a
transition redshift $(z_{tr}$. From the figure, the value of the transition
redshift for $n=2.5,{\;}2.6,{\;}2.7$ are respectively $z_{tr}=0.582,{\;}%
0.624,{\;}0.812$. The transition redshift value $z_{tr}=0.582$ at $n=2.50$
is therefore consistent with the results of the observation \cite{q1, q2, q3}%
. Hence here we fix $n=2.50$ through-out analysis.

\section{Cosmological model with $f\left( Q,T\right) =\protect\alpha Q^{m+1}+%
\protect\lambda T$}

\label{sec3}

In the first choice, we consider the cosmic evolution for a function of the
form $f\left( Q,T\right) =\alpha Q^{m+1}+\lambda T$, where $\alpha $, $%
\lambda $ and $m$ are free model parameters. In this case, we will consider
that $m\neq 0$. Hence, the expressions of $f_{Q}$ and $f_{T}$\ in the field
equation (\ref{eqn11}) are derived as $F=f_{Q}=\alpha \left( m+1\right)
Q^{m} $ and $8\pi \widetilde{G}=f_{T}=\lambda $. By using Eqs (\ref{eqn21})
with this choice, and solving the field equations (\ref{eqn17}) and (\ref%
{eqn18}), the values of $\rho $\ and $\overline{p}$ are obtained as 
\begin{widetext}
\begin{equation}
\rho =\frac{\alpha 2^{m-1}3^{m}(2m+1)}{(\lambda +4\pi )(\lambda +8\pi )}%
\left[ \lambda (m+1)\left( -\frac{\eta }{nt^{2}}\right) -3(\lambda +8\pi
)\left( \frac{\beta t+\eta }{nt}\right) ^{2}\right] \left( \frac{\beta
t+\eta }{nt}\right) ^{2m},  \label{eqn24}
\end{equation}
\begin{equation}
\overline{p}=\frac{\alpha 2^{m-1}3^{m}(2m+1)}{(\lambda +4\pi )(\lambda +8\pi
)}\left[ (3\lambda +16\pi )(m+1)\left( -\frac{\eta }{nt^{2}}\right) +3\left( 
\frac{\beta t+\eta }{nt}\right) ^{2}(\lambda +8\pi )\right] \left( \frac{%
\beta t+\eta }{nt}\right) ^{2m}.  \label{eqn25}
\end{equation}
\end{widetext}

\begin{figure}[h]
\centerline{\includegraphics[scale=0.65]{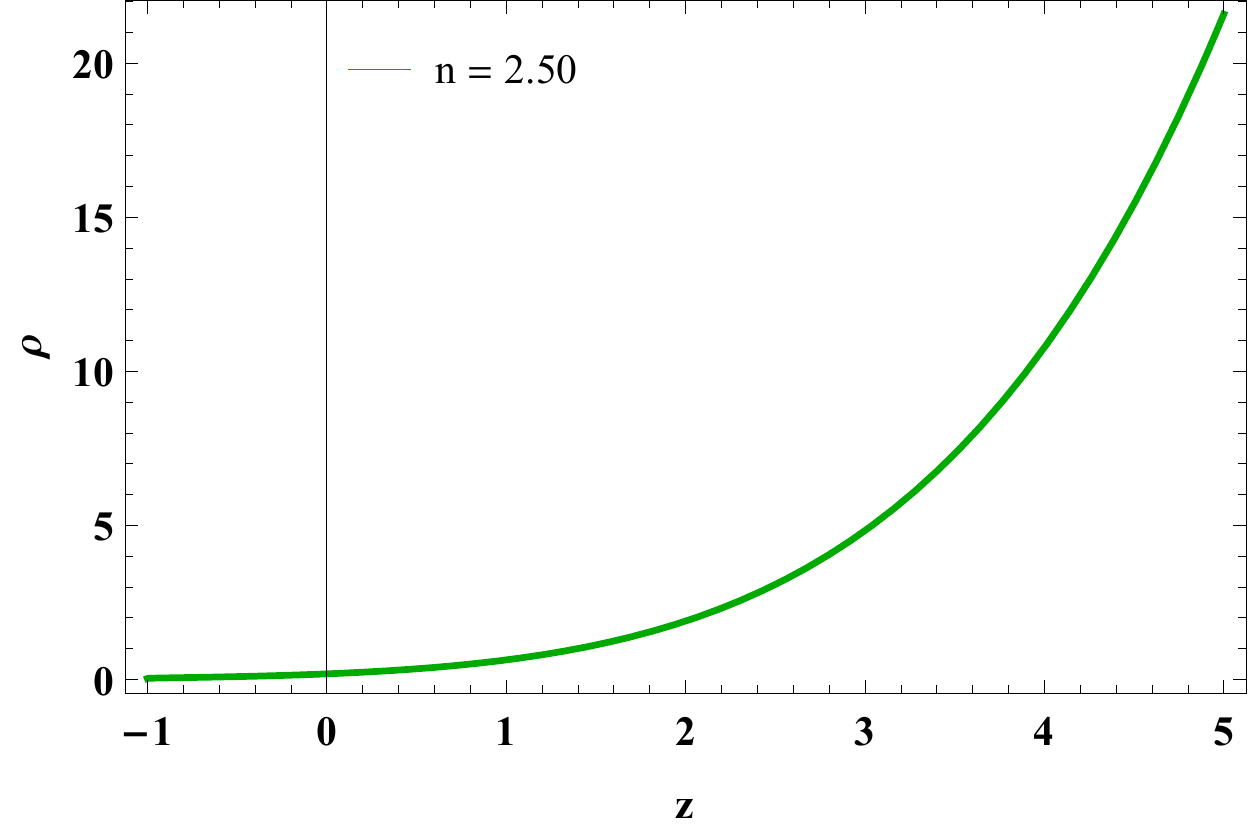}}
\caption{$\protect\rho $ versus redshift $z$ for the model $f\left(
Q,T\right) =\protect\alpha Q^{m+1}+\protect\lambda T$.}
\label{fig_rho1}
\end{figure}

\begin{figure}[h]
\centerline{\includegraphics[scale=0.65]{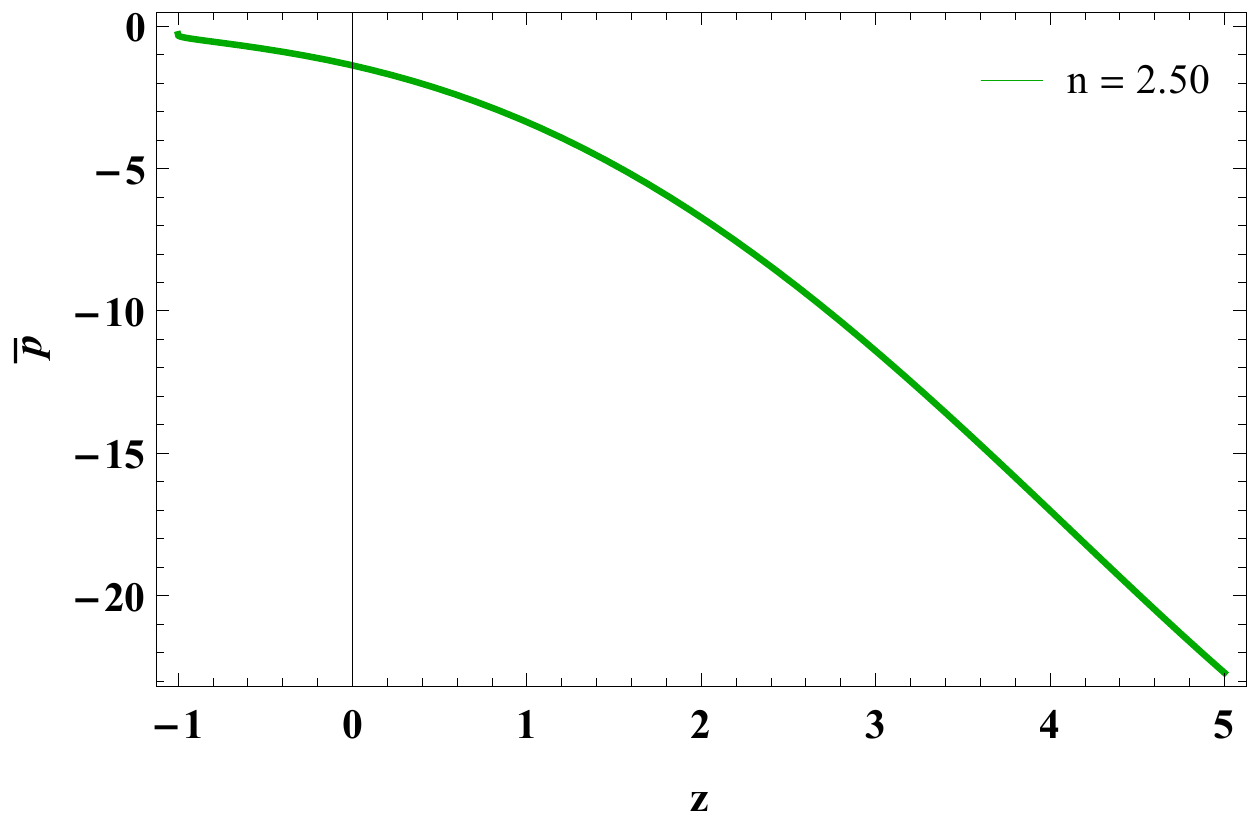}}
\caption{$\overline{p}$ versus redshift $z$ for the model $f\left(
Q,T\right) =\protect\alpha Q^{m+1}+\protect\lambda T$.}
\label{fig_p1}
\end{figure}

Fig. \ref{fig_rho1} represents the evolution of the energy density $\rho $
of the Universe in terms of the redshift $z$ for the fixed values of $\alpha
=-4.5$\textbf{, }$\lambda =-3\pi $\textbf{, }$\gamma =0.5$\textbf{\ }and%
\textbf{\ }$m=0.2$, it appears that it is an increasing function of redshift
and also remains positive for all values of $z$, i.e. throughout cosmic
evolution. Initially $\rho $ starts with a large positive value and
approaches to zero i.e. $\rho \rightarrow 0$ at $z\rightarrow 0\text{ and}-1$%
. Further, the bulk viscous pressure $\overline{p}$ behavior in terms of
redshift $z$ is shown in Fig. \ref{fig_p1}, we observe that the bulk viscous
pressure of this model takes large positive values in the beginning and
tends to zero i.e. $\overline{p}\rightarrow 0$ at the present $z=0$ and the
future $z\rightarrow -1$. According to recent observations of SNIa, the
Universe has entered an accelerating phase with negative pressure. Thus,
negative pressure characterizes the phenomenon of cosmic acceleration in the
context of modified gravity.

\begin{figure}[h]
\centerline{\includegraphics[scale=0.65]{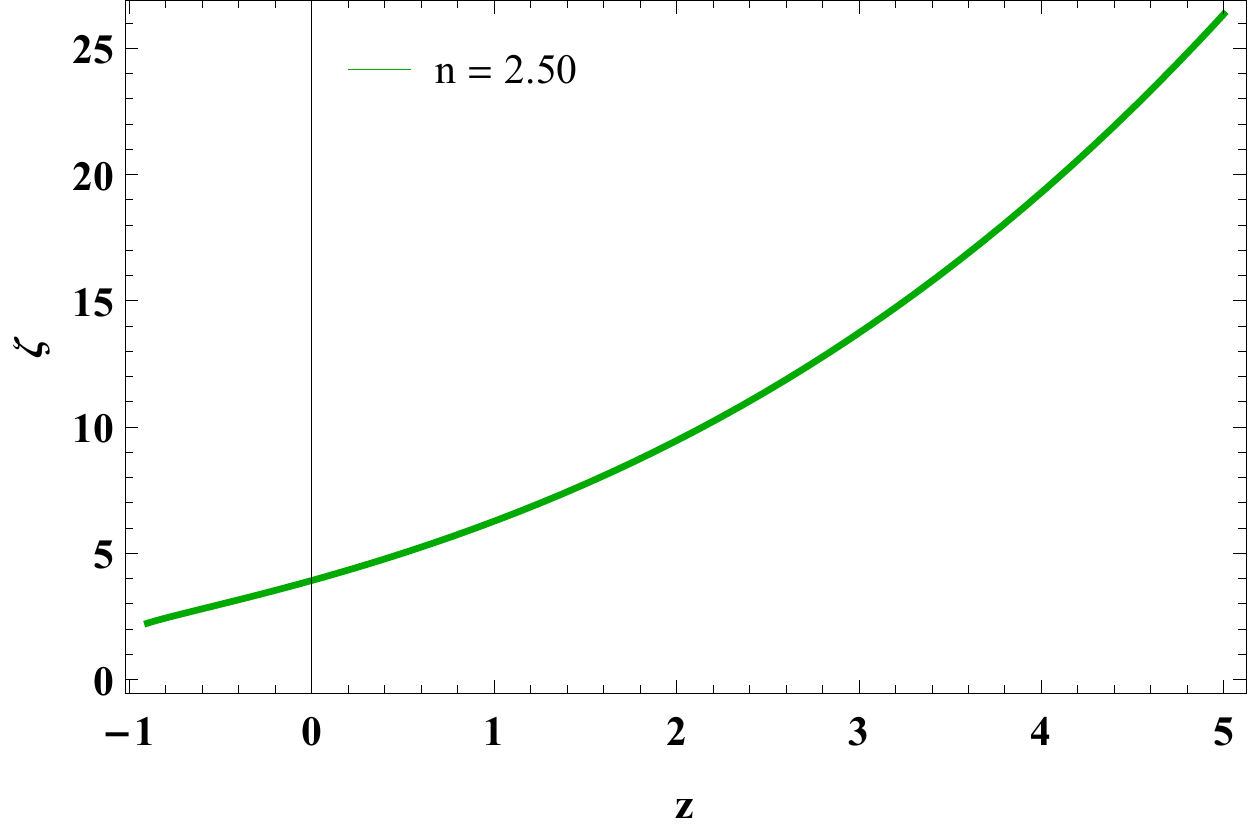}}
\caption{$\protect\xi $ versus redshift $z$ for the model $f\left(
Q,T\right) =\protect\alpha Q^{m+1}+\protect\lambda T$.}
\label{fig_xi1}
\end{figure}

\begin{figure}[h]
\centerline{\includegraphics[scale=0.65]{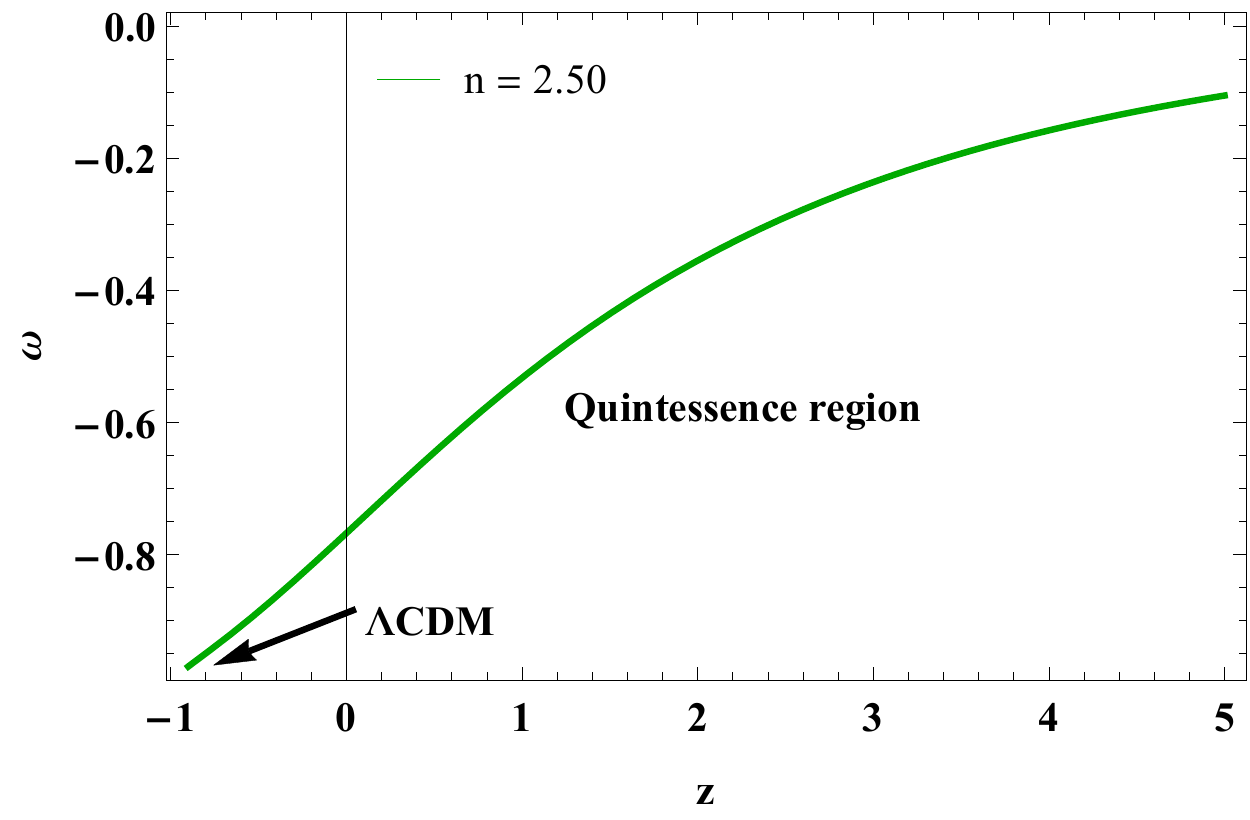}}
\caption{$\protect\omega $ versus redshift $z$ for the model $f\left(
Q,T\right) =\protect\alpha Q^{m+1}+\protect\lambda T$.}
\label{fig_EoS1}
\end{figure}

The bulk viscosity coefficient $\xi $\ and the normal pressure $p$ are
obtained as 
\begin{widetext}
\begin{equation}
\xi =-\frac{\alpha 6^{m-1}(2m+1)}{(\lambda +4\pi )(\lambda +8\pi )}\left[
\left( -\frac{\eta }{nt^{2}}\right) (m+1)(16\pi -(\gamma -3)\lambda
)+3(\gamma +1)(\lambda +8\pi )\left( \frac{\beta t+\eta }{nt}\right) ^{2}%
\right] \left( \frac{\beta t+\eta }{nt}\right) ^{2m-1},  \label{eqn26}
\end{equation}

\begin{equation}
p=\gamma \rho =\frac{\alpha \gamma 2^{m-1}3^{m}(2m+1)}{(\lambda +4\pi
)(\lambda +8\pi )}\left[ \lambda (m+1)\left( -\frac{\eta }{nt^{2}}\right)
-3(\lambda +8\pi )\left( \frac{\beta t+\eta }{nt}\right) ^{2}\right] \left( 
\frac{\beta t+\eta }{nt}\right) ^{2m}.  \label{eqn27}
\end{equation}
\end{widetext}In thermodynamics, the behavior of the bulk viscosity
coefficient $\xi $ is positive, ensuring that viscosity pushes the bulk
viscous pressure toward negative values. From Fig. \ref{fig_xi1}, it is
clear that the bulk viscosity coefficient $\xi $ is an increasing function
of redshift $z$, remains positive for all values of $z$, and approaches a
constant amount close to zero, which resumbles well with the physical
behavior of $\xi $ \cite{ref14}.

The EoS parameter $\omega $ is given by

\begin{equation}
\omega=\frac{(3\lambda +16\pi )(m+1)\left( -\frac{\eta }{nt^{2}}\right)
+3\left( \frac{\beta t+\eta }{nt}\right) ^{2}(\lambda +8\pi )}{\lambda
(m+1)\left( -\frac{\eta }{nt^{2}}\right) -3(\lambda +8\pi )\left( \frac{%
\beta t+\eta }{nt}\right) ^{2}}.  \label{eqn28}
\end{equation}

The EoS parameter is a function of energy density and pressure i.e. $\omega =%
\frac{\overline{p}}{\rho }$. Recent works published in the literature
confirm that the value of the EoS parameter is currently in the specified
range $-1\leq \omega \leq 0$ due to the acceleration phase of the Universe.
In the case where the value of $\omega =-1$, we say that the model is a
cosmological constant $\left( \Lambda \right) $. Further, in the case of $%
-1<\omega \leq -\frac{1}{3}$ we say we have a quintessence model. Finally,
if $\omega <-1$ the model is like a phantom model. Also, the several
cosmological analysis have constrained the numerical value of the EoS
parameter such as: Supernovae Cosmology Project, $\omega = -1.035^{+0.055}_{
-0.059}$ \cite{R}; WMAP+CMB, $\omega = -1.073^{+0.090}_{ -0.089}$ \cite{G};
Planck 2018, $\omega = -1.03 \pm 0.03$ \cite{N}.

Here, the behavior of EoS parameter $\omega $\ in terms of
redshift $z$\ is shown in Fig. \ref{fig_EoS1}. The EoS parameter
indicates that the bulk viscous fluid behaves like the quintessence dark
energy model at $z=0$\ and is finally approached to $\Lambda $%
CDM region at $z\rightarrow -1$\ which describe the
late-time acceleration of the expanding universe without invoking any dark
energy component. Further, the present value of EoS parameter corresponding
to the parameters of the model $\alpha =-4.5$, $\lambda =-3\pi $, $\gamma
=0.5$ and $m=0.2$\ is $\omega _{0}=-0.7698$. Thus, the
cosmic fluid with bulk viscosity is the most viable candidate and thus the
EoS parameter of the model is in good agreement with the astronomical
observations \cite{ref4}.

As we have the cosmological parameters studied above play an important role
in understanding the evolution of the Universe. But to predict the cosmic
acceleration in modern cosmology, a set of energy conditions can be obtained
from the equation of Raychaudhuri. In GR, the role of these energy
conditions is to demonstrate the theorems for the existence of space-time
singularity and black holes \cite{ref43}. Various authors have worked on
energy conditions in different backgrounds. In this work, we will consider
the renowned energy conditions to test the validity of the model in the
context of cosmic acceleration. There are several forms of energy conditions
such as weak energy conditions (WEC), null energy conditions (NEC), dominant
energy conditions (DEC), and strong energy conditions (SEC) are given for
the content of the Universe in form of a viscous fluid in $f\left(
Q,T\right) $ gravity as follows \cite{ref44}

\begin{itemize}
\item WEC: if $\rho \geq 0,\rho +\overline{p}\geq 0$;

\item NEC: if $\rho +\overline{p}\geq 0$;

\item DEC: if $\rho \geq 0,\left\vert \overline{p}\right\vert \leq \rho $;

\item SEC: if $\rho +3\overline{p}\geq 0$.
\end{itemize}

\begin{figure}[h]
\centerline{\includegraphics[scale=0.61]{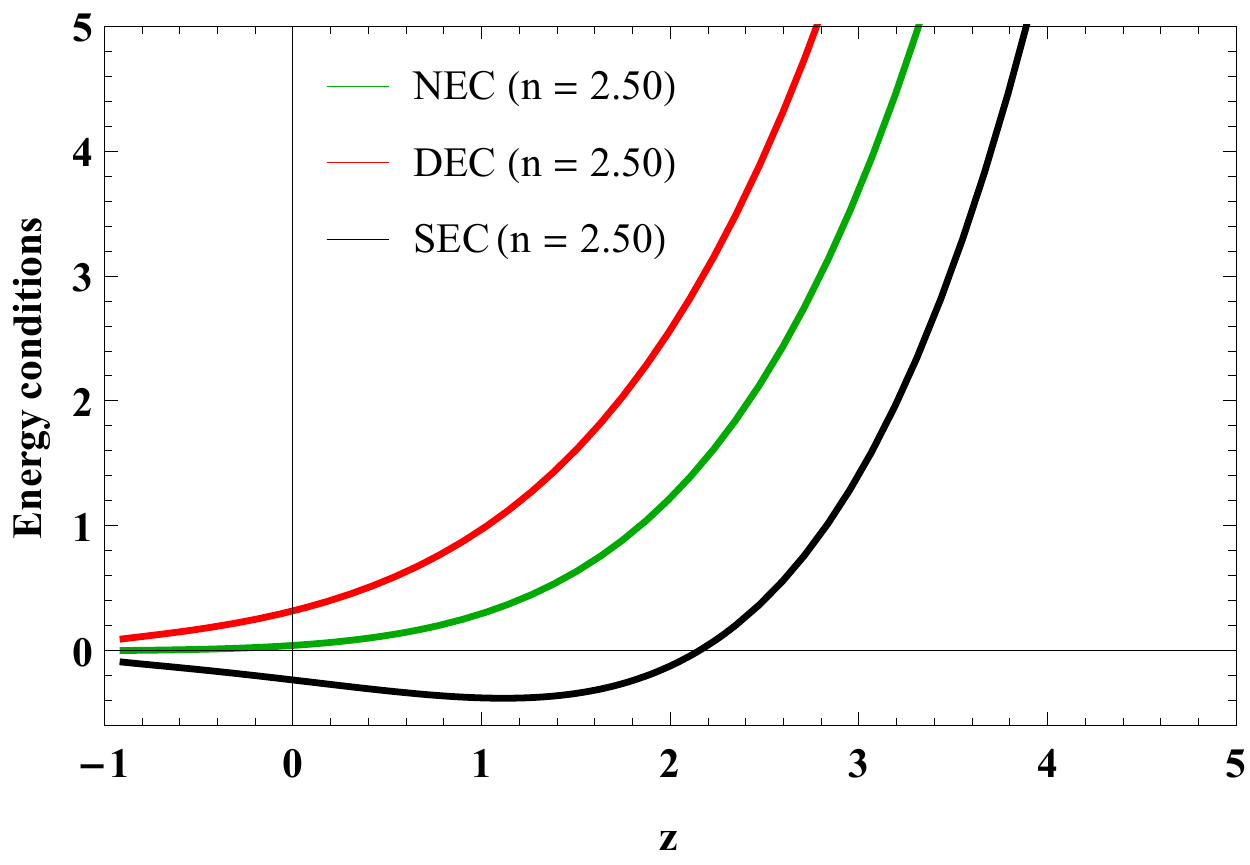}}
\caption{NEC, DEC and SEC versus redshift $z$ for the model $f\left(
Q,T\right) =\protect\alpha Q^{m+1}+\protect\lambda T$.}
\label{fig_NEC1}
\end{figure}

The importance of the above energy conditions is that when the NEC is
violated, all more energy conditions are violated. This violation of NEC
constitutes the depletion of energy density as the Universe expands.
Further, the violation of SEC constitutes the acceleration of the Universe.
To account for the late-time cosmic acceleration with $\omega \simeq -1 $,
it must be $\rho \left( 1+3\omega \right) <0$. Using Eqs. (\ref{eqn24}) and (%
\ref{eqn25}) we can get the behavior of energy conditions in terms of $z $
and $n$ of the model as shown in Fig \ref{fig_NEC1}. We observe that WEC,
NEC and DEC are satisfied while the SEC is violated in the present and
future. Hence, the violation of SEC leads to the acceleration of the
Universe.

\begin{figure}[h]
\centerline{\includegraphics[scale=0.65]{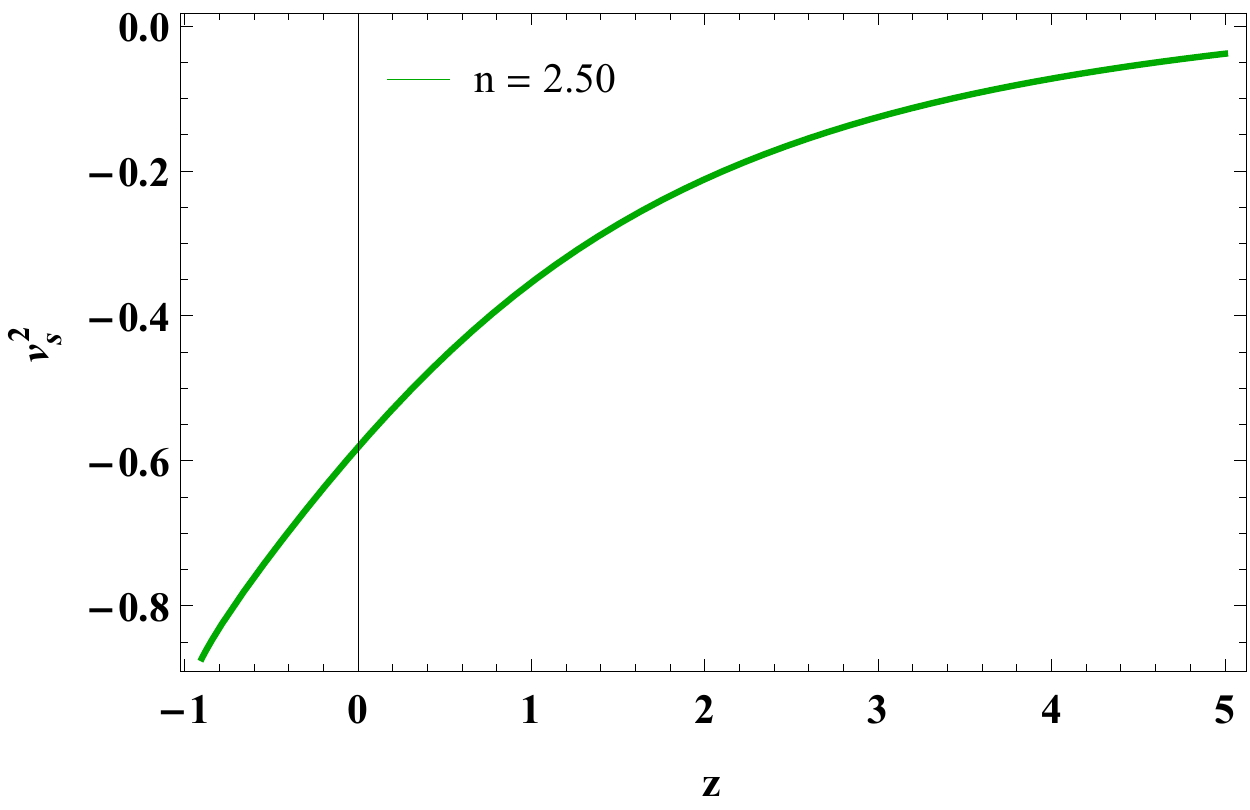}}
\caption{$\protect\vartheta^{2}_{s}$ versus redshift $z$ for the model $%
f\left( Q,T\right) =\protect\alpha Q^{m+1}+\protect\lambda T$.}
\label{fig_st_1}
\end{figure}

The stability parameter As, there are three types of particles are available
in the universe called sub-luminal, luminal and super-luminal. Out of which
the sub-luminal particles moving very slow in comparison of the speed of
light (example are electrons and neutrons while the luminal particles move
with exactly the same speed as that of the speed of light (example are
photon and graviton) whereas super-luminal particles are moving faster than
the speed of light (tachyons). There are two possibilities for the existence
of super-luminal particles: either they do not exist or if they do, then
they do not interact with an ordinary matter. If the speed of sound is less
than the local light speed, then only we can say about the non-violation of
causality. The positive square sound speed is necessary for the classical
stability of the universe. The speed of sound $\vartheta _{s}^{2}$ is
defined by $\vartheta _{s}^{2}=\frac{\partial p}{\partial \rho }$ and
observed as 
\begin{equation}
\vartheta _{s}^{2}=\frac{(3\lambda +16\pi )n(\eta +\eta m+\beta t)-3(\lambda
+8\pi )(\eta +\beta t)^{2}}{\lambda \left( n(\eta +\eta m+\beta t)+3(\eta
+\beta t)^{2}\right) +24\pi (\eta +\beta t)^{2}}.
\end{equation}%
The Fig. \ref{fig_st_1}, shows that the stability parameter $\vartheta
_{s}^{2}<0$ throughout the evolution of the universe. Hence, the model
remains unstable with the expansion of the Universe.

\section{Cosmological model with $f\left( Q,T\right) =\protect\alpha Q+%
\protect\lambda T$}

\label{sec4}

\begin{figure}[h]
\centerline{\includegraphics[scale=0.65]{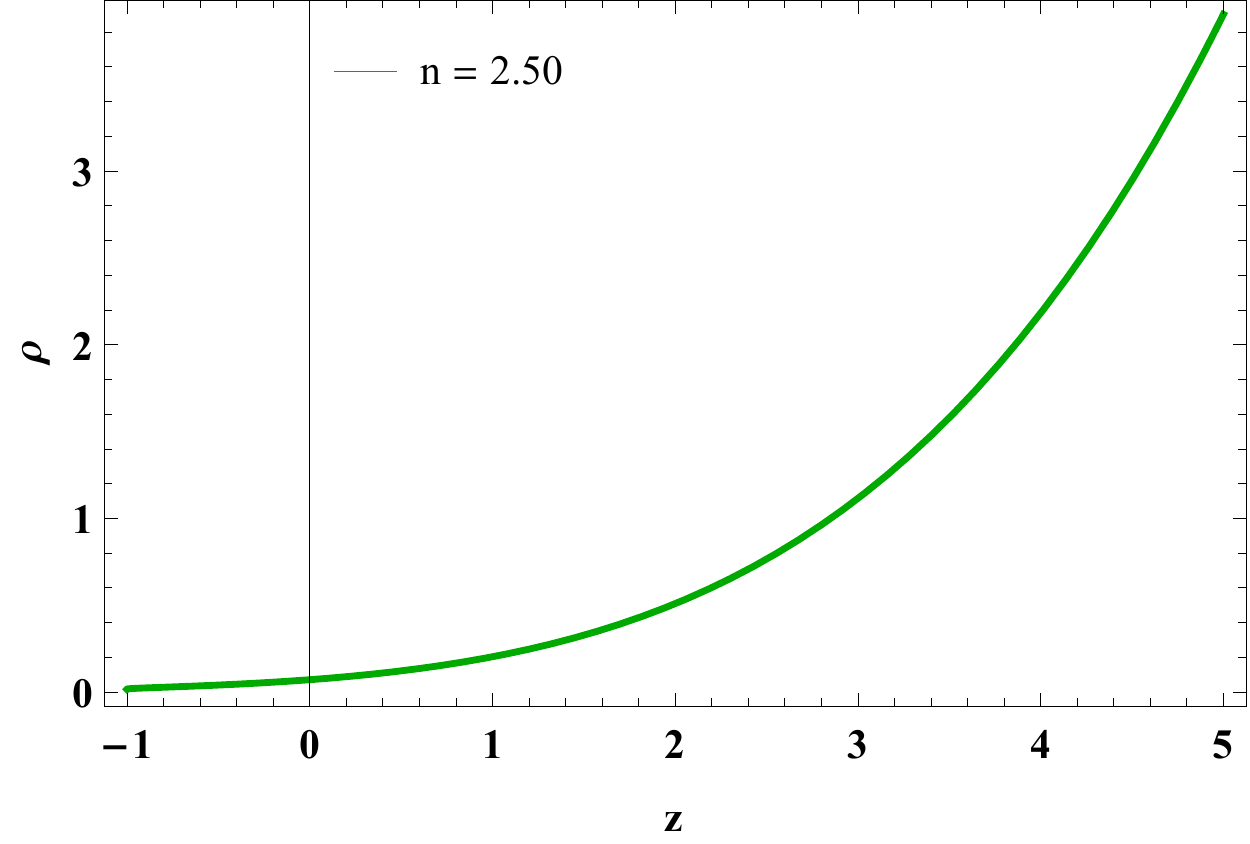}}
\caption{$\protect\rho $ versus redshift $z$ for the model $f\left(
Q,T\right) =\protect\alpha Q+\protect\lambda T$.}
\label{fig_rho2}
\end{figure}

\begin{figure}[h]
\centerline{\includegraphics[scale=0.65]{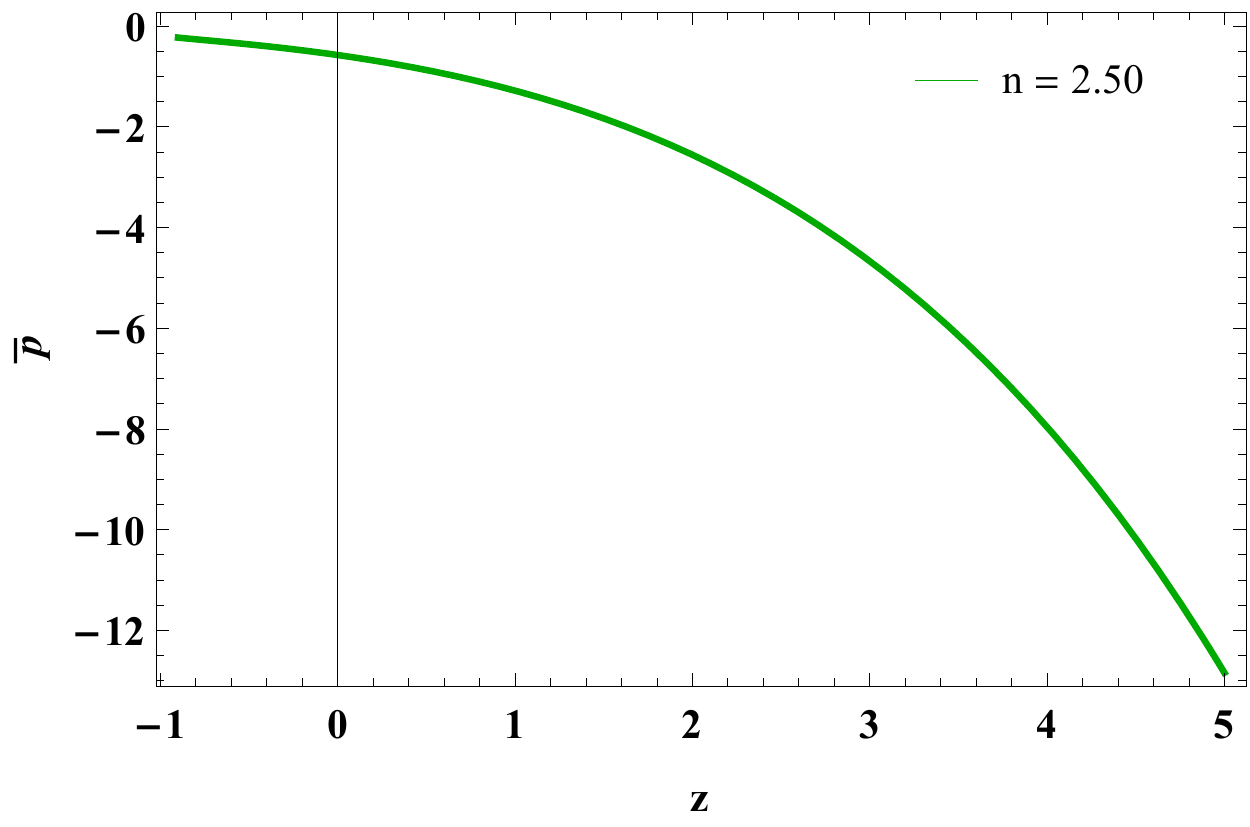}}
\caption{$\overline{p}$ versus redshift $z$ for the model $f\left(
Q,T\right) =\protect\alpha Q+\protect\lambda T$.}
\label{fig_p2}
\end{figure}

For the second model, we presume a functional form of $f\left( Q,T\right) $
as $f\left( Q,T\right) =\alpha Q+\lambda T$, where $\alpha $ and $\lambda $
are free model parameters ($m=0$). In this case, we get $F=f_{Q}=\alpha $
and $8\pi \widetilde{G}=f_{T}=\lambda $. Here, using Eqs. (\ref{eqn19}) with
this case, and solving the field equations (\ref{eqn17}) and (\ref{eqn18}),
the values of $\rho $\ and $\overline{p}$ are obtained as 
\begin{widetext}
\begin{equation}
\rho =\frac{\alpha }{2(\lambda +4\pi )(\lambda +8\pi )}\left[ \lambda \left(
-\frac{\eta }{nt^{2}}\right) -3(\lambda +8\pi )\left( \frac{\beta t+\eta }{nt%
}\right) ^{2}\right],  \label{eqn32}
\end{equation}

\begin{equation}
\overline{p}=\frac{\alpha }{2(\lambda +4\pi )(\lambda +8\pi )}\left[
(3\lambda +16\pi )\left( -\frac{\eta }{nt^{2}}\right) +3(\lambda +8\pi
)\left( \frac{\beta t+\eta }{nt}\right) ^{2}\right].  \label{eqn33}
\end{equation}
\end{widetext}Fig. \ref{fig_rho2} represents the evolution of the energy
density $\rho $ of the Universe in terms of the redshift $z$ for the fixed $%
\alpha =-4.5$, $\lambda =-3\pi $ and $\gamma =0.5$. We can observe that the
energy density is an increasing function of redshift and remains positive.
Initially, $\rho $ starts with a large positive value and approaches zero at 
$z\rightarrow -1$. The bulk viscous pressure $\overline{p}$ behavior in
terms of redshift $z$ is shown in Fig. \ref{fig_p2}. We can see that the
bulk viscous pressure of this model is initially positive and then becomes
negative in the present and the future, and it is an increasing function of
redshift for all fixed $n,\alpha ,\lambda $.

\begin{figure}[h]
\centerline{\includegraphics[scale=0.65]{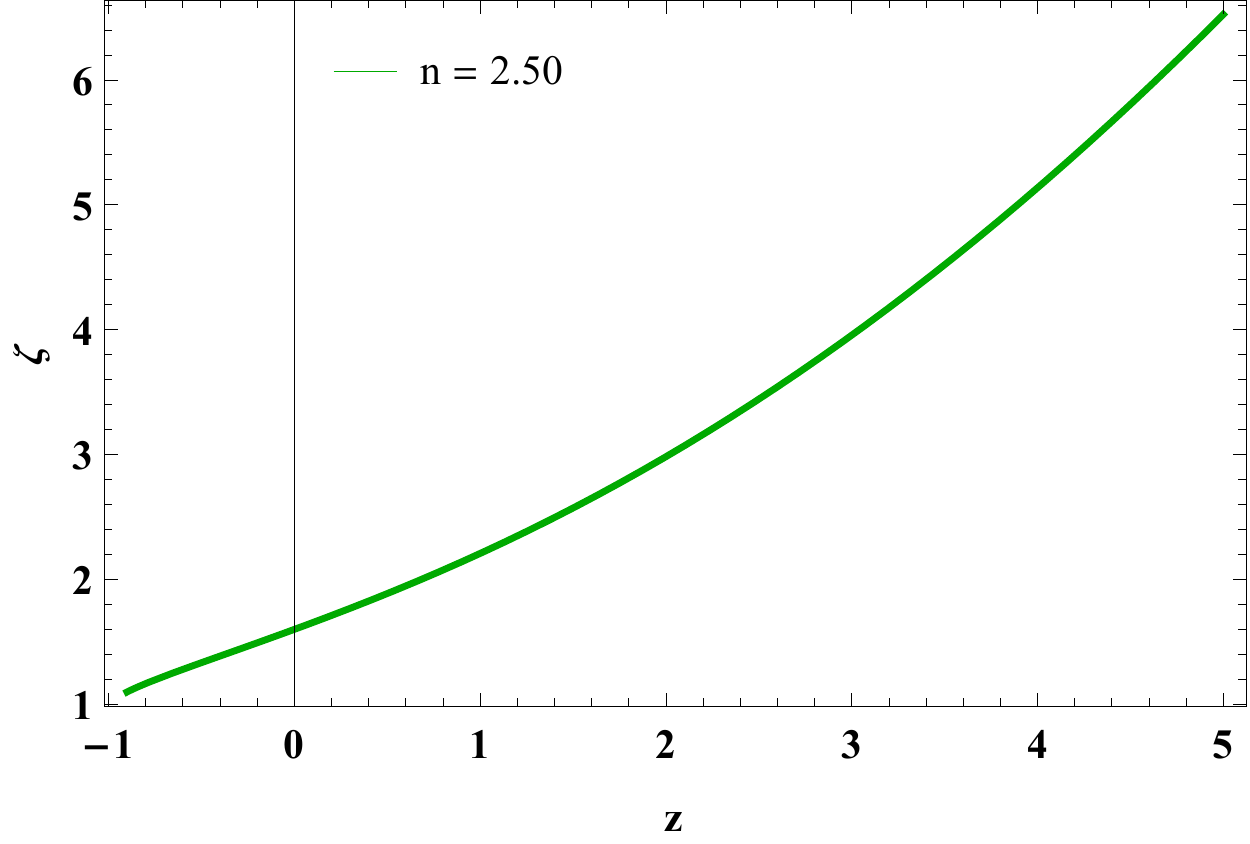}}
\caption{$\protect\xi $ versus redshift $z$ for the model $f\left(
Q,T\right) =\protect\alpha Q+\protect\lambda T$.}
\label{fig_xi2}
\end{figure}

For this choice of model $f$, the values of bulk viscosity coefficient $\xi $%
\ and normal pressure $p$ are obtained as 
\begin{widetext}
\begin{equation}
\xi=-\frac{\alpha }{6(\lambda +4\pi )(\lambda
+8\pi )}\left[ (16\pi -\lambda (\gamma -3))\left( -\frac{\eta }{nt^{2}}%
\right) +3(\gamma +1)(\lambda +8\pi )\left( \frac{\beta t+\eta }{nt}\right)
^{2}\right] \left( \frac{nt}{\beta t+\eta }\right),  \label{eqn34}
\end{equation}

\begin{equation}
p=\gamma \rho =\frac{\alpha \gamma }{2(\lambda +4\pi )(\lambda +8\pi )}\left[
\lambda \left( -\frac{\eta }{nt^{2}}\right) -3(\lambda +8\pi )\left( \frac{%
\beta t+\eta }{nt}\right) ^{2}\right].  \label{eqn35}
\end{equation}
\end{widetext}Through Fig. \ref{fig_xi2}, it is clear that the bulk
viscosity coefficient $\xi $ is an increasing function of redshift $z$,
remains positive for all values of $z$ i.e. throughout the evolution of the
Universe.

\begin{figure}[h]
\centerline{\includegraphics[scale=0.65]{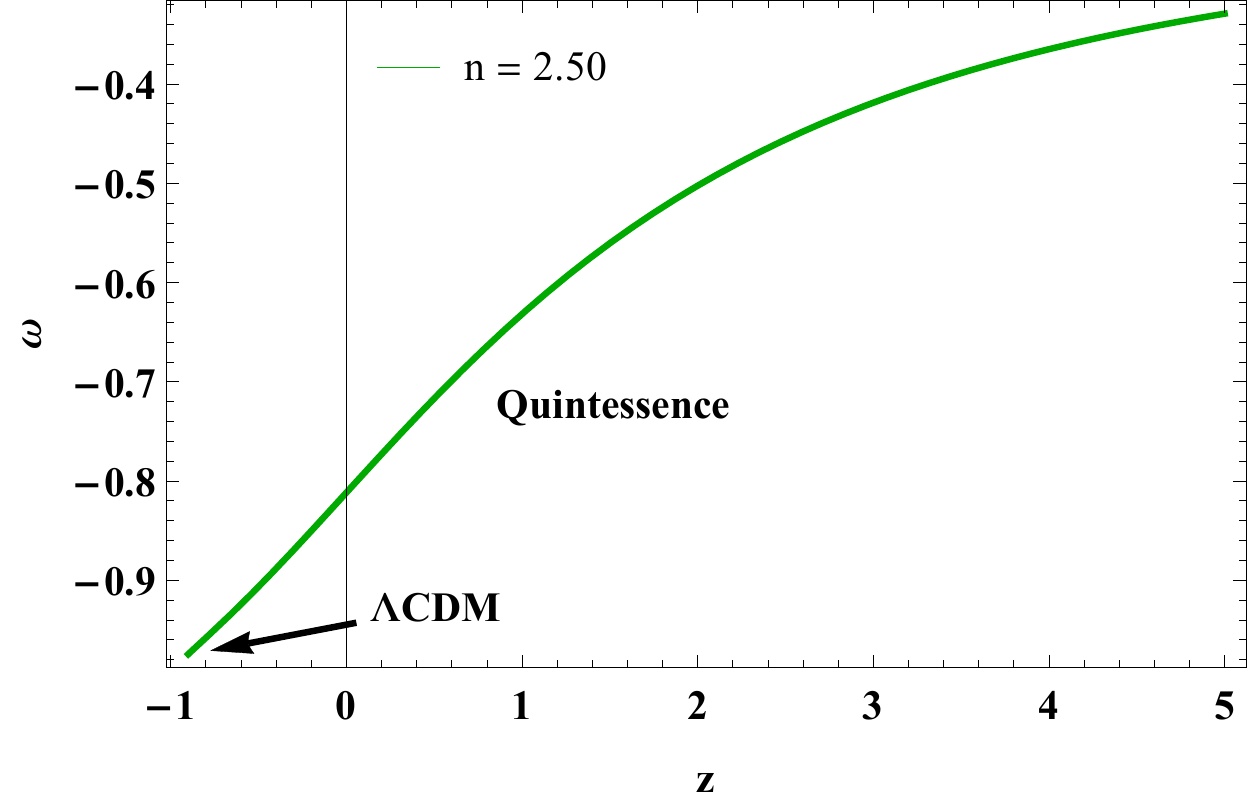}}
\caption{$\protect\omega $ versus redshift $z$ for the model $f\left(
Q,T\right) =\protect\alpha Q+\protect\lambda T$.}
\label{fig_EoS2}
\end{figure}

The EoS parameter $\omega $ is given by

\begin{equation}
\omega=\frac{(3\lambda +16\pi )\left( -\frac{\eta }{nt^{2}}\right)
+3(\lambda +8\pi )\left( \frac{\beta t+\eta }{nt}\right) ^{2}}{\lambda
\left( -\frac{\eta }{nt^{2}}\right) -3(\lambda +8\pi )\left( \frac{\beta
t+\eta }{nt}\right) ^{2}}.  \label{eqn36}
\end{equation}

\begin{figure}[h]
\centerline{\includegraphics[scale=0.62]{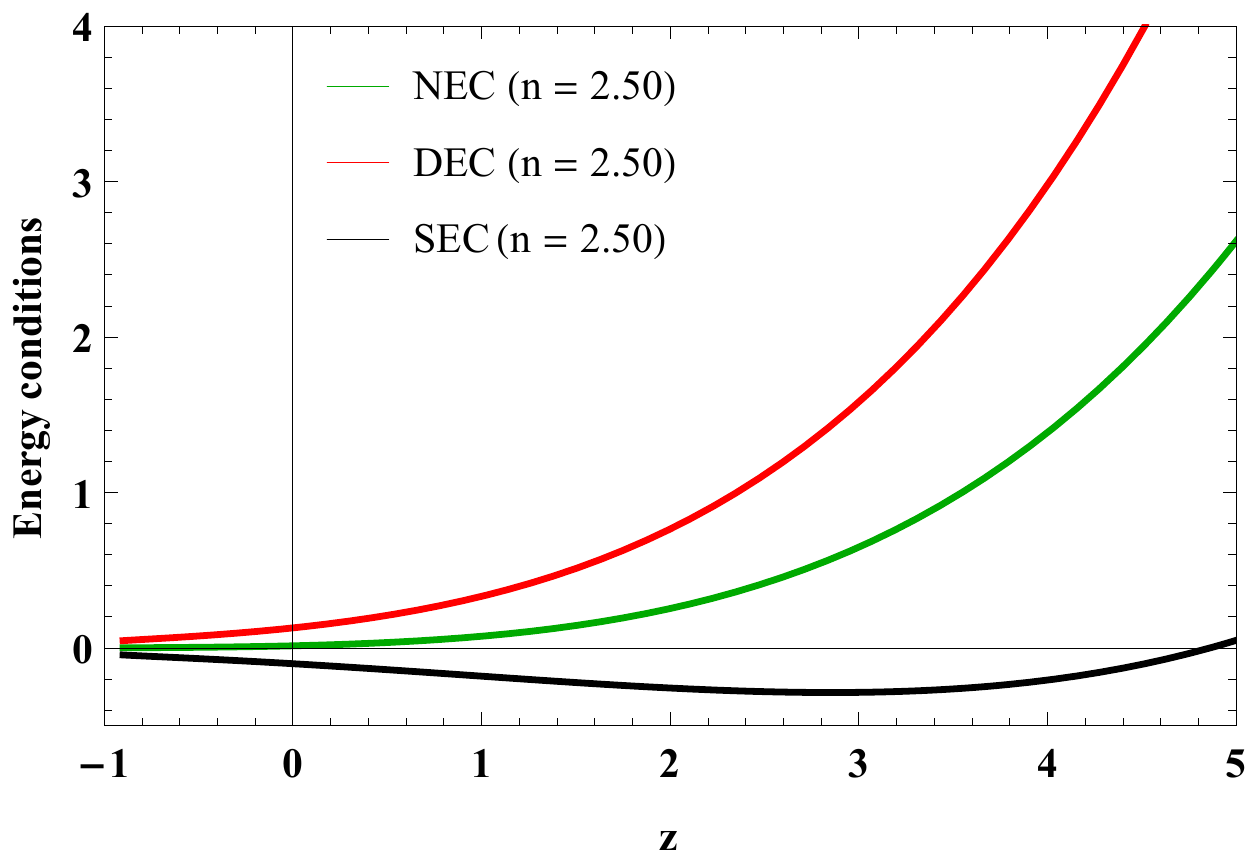}}
\caption{NEC, DEC and SEC versus redshift $z$ for the model $f\left(
Q,T\right) =\protect\alpha Q+\protect\lambda T$.}
\label{fig_NEC2}
\end{figure}

For this case, the variation of EoS parameter $\omega $ in terms of redshift 
$z$ is shown in Fig. \ref{fig_EoS2}. It can be seen that the bulk
viscous fluid behaves like the quintessence dark energy model at $z=0$%
\ and is finally approached to $\Lambda $CDM region at $%
z\rightarrow -1$\ like the first case. In addition, the present
value of EoS parameter corresponding to the parameters of the model $\alpha
=-4.5$, $\lambda =-3\pi $\ and $\gamma =0.5$\ is $%
\omega _{0}=-0.8122$. Thus, the EoS parameter for this case is in
good agreement with the astronomical observations. From Fig. \ref{fig_NEC2}%
, we observe that WEC, NEC and DEC are satisfied while the SEC is violated
in the present and future. 
The stability parameter

The speed of sound $\vartheta^{2}_{s}$ is observed as

\begin{equation}
\vartheta^{2}_{s}=\frac{(3 \lambda +16 \pi ) n-3 (\lambda +8 \pi ) (\eta
+\beta t)}{\lambda (n+3 (\eta +\beta t))+24 \pi (\eta +\beta t)}.
\end{equation}

\begin{figure}[h]
\centerline{\includegraphics[scale=0.65]{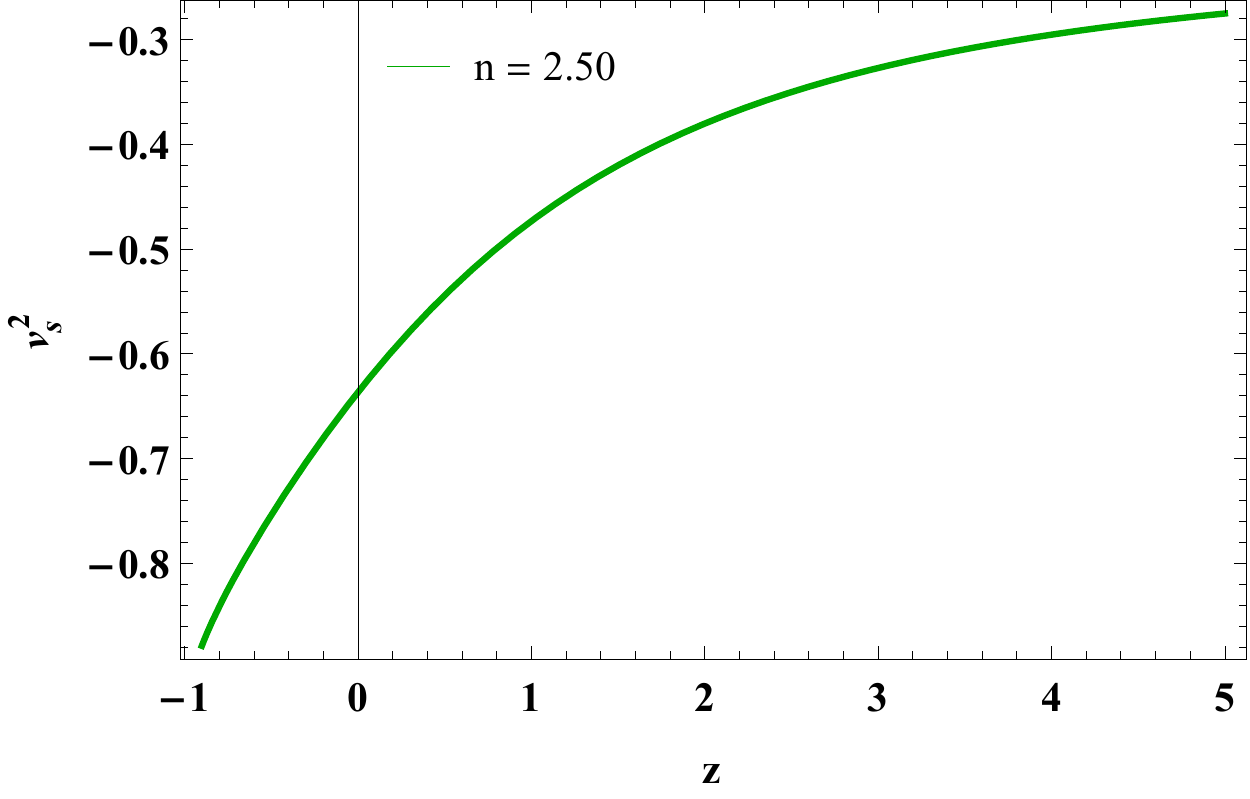}}
\caption{$\protect\vartheta^{2}_{s}$ versus redshift $z$ for the model $%
f\left( Q,T\right) =\protect\alpha Q+\protect\lambda T$.}
\label{fig_st_2}
\end{figure}

The Fig. \ref{fig_st_2}, shows that the stability parameter $\vartheta
_{s}^{2}<0$ throughout the evolution of the universe. Thus, the model
remains unstable with the expansion of the Universe.

\section{Discussion and conclusion}

\label{sec5}

In this paper, we investigated the flat FLRW cosmological models in
existence of bulk viscosity in the framework of extended symmetric
teleparallel gravity. We considered two $f\left( Q,T\right) $ models,
specifically, $f\left( Q,T\right) =\alpha Q^{m+1}+\lambda T$ and $f\left(
Q,T\right) =\alpha Q+\lambda T$ where $\alpha $, $\lambda $ and $m$ are free
models parameters. To find the exact solutions of the field equations we
used the hybrid expansion law. According to the relevant recent
observational data, the expansion phase of the Universe is accelerating,
which means that the current deceleration parameter should be in the range
of $-1\leq q\leq 0$. Using the assumed redshift-time relation and the hybrid
scale factor, the behavior of the deceleration parameter was plotted in
terms of redshift in Fig. \ref{fig_q}. Knowing that the parameters of the
model $\eta $, $\beta $ and $n$ have their values carefully chosen to comply
with the observational constraints i.e. $\eta =1.35$, $\beta =2.05$ and
three values for $n=2.5$, $2.6$, $2.7$. Based on these data, we found that
the deceleration parameter for our model is positive in the early phase of
the Universe and negative for the present and late Universe. It indicates
that the Universe shows a transition from deceleration to acceleration at
the transition redshift value $z_{tr}=0.582$ for $n=2.50$ with current
values of the deceleration parameter for our model within the range of the
observational data.

From Figs. \ref{fig_rho1} and \ref{fig_rho2}, we observed that for both
models, the energy density is an increasing function of redshift and remains
positive. It starts with a large positive value and approaches to zero at $%
z\rightarrow -1$. The behavior of bulk viscous pressure for $m\neq 0$ and $%
m=0$ is represented in Figs. \ref{fig_p1} and \ref{fig_p2} respectively, and
we see that it is negative in the present and the future which is consistent
with recent observations. Further, the behavior of the bulk viscosity
coefficient for both the models is a positive and an increasing function of
redshift $z$. Thus, this behavior is consistent with thermodynamics. The
variation of EoS parameter versus redshift for both models is shown in Figs. %
\ref{fig_EoS1} and \ref{fig_EoS2} respectively. In both models, the EoS
parameter behaves like the quintessence model in the present and the
cosmological constant model in the future. Finally, from the energy
conditions we can conclude that WEC, NEC and DEC are satisfied while the SEC
is violated in the present and future. Also, both models are unstable
throughout the evolution of the universe. The physical behavior of all the
cosmological parameters discussed in this work of both models is remarkably
consistent with the observational data.\newline

\section*{Acknowledgments}

We are very much grateful to the honorary referee and the editor for the
illuminating suggestions that have significantly improved our work in terms
of research quality and presentation.

\textbf{Data availability} There are no new data associated with this
article.

\textbf{Declaration of competing interest} The authors declare that they
have no known competing financial interests or personal relationships that
could have appeared to influence the work reported in this paper.\newline

\end{document}